\documentclass{moriond}

\def\be{\begin{equation}}
\def\ee{\end{equation}}
\def\bea{\begin{eqnarray}}
\def\eea{\end{eqnarray}}

\usepackage{wrapfig,rotating}
\usepackage{enumerate}

\newcommand{\Photo}{\includegraphics[height=35mm]{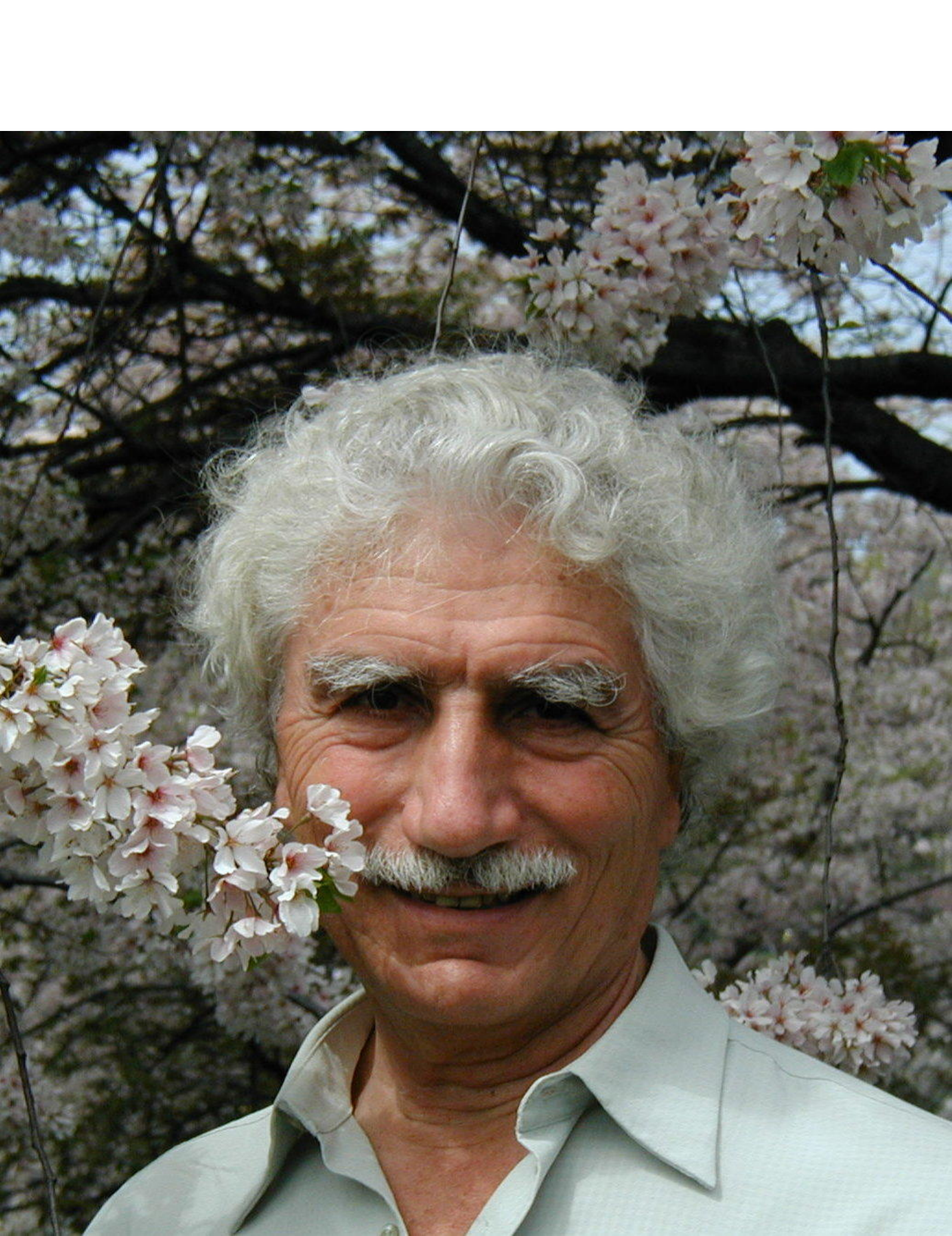}}

\begin{document}
\vspace*{4cm}
\title{RENORM tensor-Pomeron diffractive predictions}

\author{K. Goulianos}

\address{The Rockefeller University, 1230 York Avenue, New York, NY 10065-6399, USA}\footnote{Presented at {\em Moriond QCD and High Energy Interactions,} La Thuile, Italy, March 22-29, 2016.}

\maketitle\abstracts{
Predictions of the elastic scattering, total-inetastic, and total proton-proton cross sections, based on a Regge theory inspired tensor-Pomeron implementation of the RENORM model for hadronic diffraction, are compared to the latest experimental measurements at the LHC. The measured cross sections are in good agreement within the experimental uncertainties of the data and the theoretical uncertainties of the model, reaching down to the ~1\% level.}

\section{Introduction}
 In {\sc dis-2015} (Spring 2015), we summarized~\cite{ref:dis2015_paper}, the pre-{\sc lhc} predictions of the total, elastic and total-inelastic, as well as the  single- and double-diffractive components of the proton-proton cross section at high energies, based on the {\sc renorm}/{\sc mbr} model~\cite{ref:dinoModel}. 
We compared the measurements of the {\sc sd} and {\sc dd} cross sections from the Tevatron and the {\sc lhc} with the predictions of the model and found excellent agreement. 
Good agreement was also observed between the model predictions and the total, elastic, and total inelastic cross sections obtained at the Tevatron at $\sqrt s=1.8$~TeV, and at the {\sc lhc} at $\sqrt s=$~7 and 8~TeV. 

The success of the predictions of the {\sc renorm}/{\sc mbr} model for all the above cross sections  at the Tevatron and {\sc lhc} up to $\sqrt s=8$~TeV prompted an extrapolation to $\sqrt s=13$~TeV, the nominal foreseen colliding-beam energy at the {\sc lhc} in Summer 2015. For $\sigma_{\rm tot}$, $\sigma_{\rm el}$ and $\sigma_{\rm inel}$, we predicted 108~mb, 32~mb and 77~mb, respectively, with uncertainties of $\sim 11$\% in all cases, mainly due to the uncertainty in the energy-squared scale parameter $s_0$ of the model.

In Summer 2015, we updated the value of $s_0$ to a more precise one based  on a tensor glueball interpretation of the Axial Field Spectrometer (AFS) exclusive charged di-pion data~\cite{ref:AlbrowCouhlinForshaw} \cite{ref:Albrow_private} \cite{ref:PeterCesilThesis}. This change in {\sc renorm}/{\sc mbr} decreases  the uncertainties in the predictions  of the total, elastic, and total-inelastic cross sections to less than $2\%$ from Tevatron to {\sc lhc} energies, with little or no effect on the mean values, and yields cross sections in excellent agreement with the measurements by {\sc atlas} at $\sqrt s=7$~TeV and by {\sc totem} at $\sqrt s=7$ and 8~TeV, as discussed below.

\section{RENORM cross sections}
The total, elastic, and total-inelastic cross sections in the {\sc renorm/mbr} model depend on the value of the energy-squared scale parameter, $s_0$.   
Quoting verbatim from Ref.~\cite{ref:dis2015_paper},

``The total cross section ($\sigma_{\rm tot}$) is expressed as~\cite{ref:MBR_note} 
\begin{eqnarray}
\sigma^{p^{\pm}p}_{\rm tot} = &16.79 s^{0.104} + 60.81 s^{-0.32} \mp 31.68 s^{-0.54}& {\rm for}\; \sqrt{s} \le 1.8 \mbox{ TeV},\\
\sigma^{p^{\pm}p}_{\rm tot} = &\sigma_{\rm tot}^{\rm CDF}+\frac{\pi}{s_0}\left[ \left(\ln\frac{s}{s_F}\right)^2- \left(\ln\frac{s^{\rm CDF}}{s_F} \right)^2\right] & {\rm for}\; \sqrt{s} \ge 1.8 \mbox{ TeV},
\label{eqTOT}
\end{eqnarray}
where $s_0$ and $s_F$ are the energy and (Pomeron flux) saturation scales, $s_0=3.7\pm 1.5$~GeV$^2$ and $\sqrt s_F=22$~GeV, respectively. For  $\sqrt{s} \le 1.8$ TeV, where there are Reggeon contributions, we use the global fit expression~\cite{ref:CMG_96}, while for $\sqrt{s} \ge 1.8$ TeV, where Reggeon contributions are negligible,  we employ the Froissart-Martin formula~\cite{ref:Froissart,ref:Martin1966,ref:Martin2009}. The two expressions are smoothly matched at $\sqrt{s} \approx 1.8$ TeV. 
The $\sigma_{\rm el}$  for $\sqrt{s} \le 1.8$ TeV is obtained from the global fit~\cite{ref:CMG_96}, while for $1.8<\sqrt s\leq 50$ TeV we use an extrapolation of the global-fit ratio of $\sigma_{\rm el}/\sigma_{\rm tot}$, which is slowly varying with $\sqrt s,$ multiplied by $\sigma_{tot}$. The total non-diffractive cross section is given by $\sigma_{\rm ND}=(\sigma_{\rm tot} - \sigma_{\rm el}) - (2\sigma_{\rm SD}+\sigma_{\rm DD}+\sigma_{\rm CD})$.''

\section{Tensor-Pomeron predictions}
\begin{figure}[h]
\centerline{\includegraphics[width=.8\textwidth]{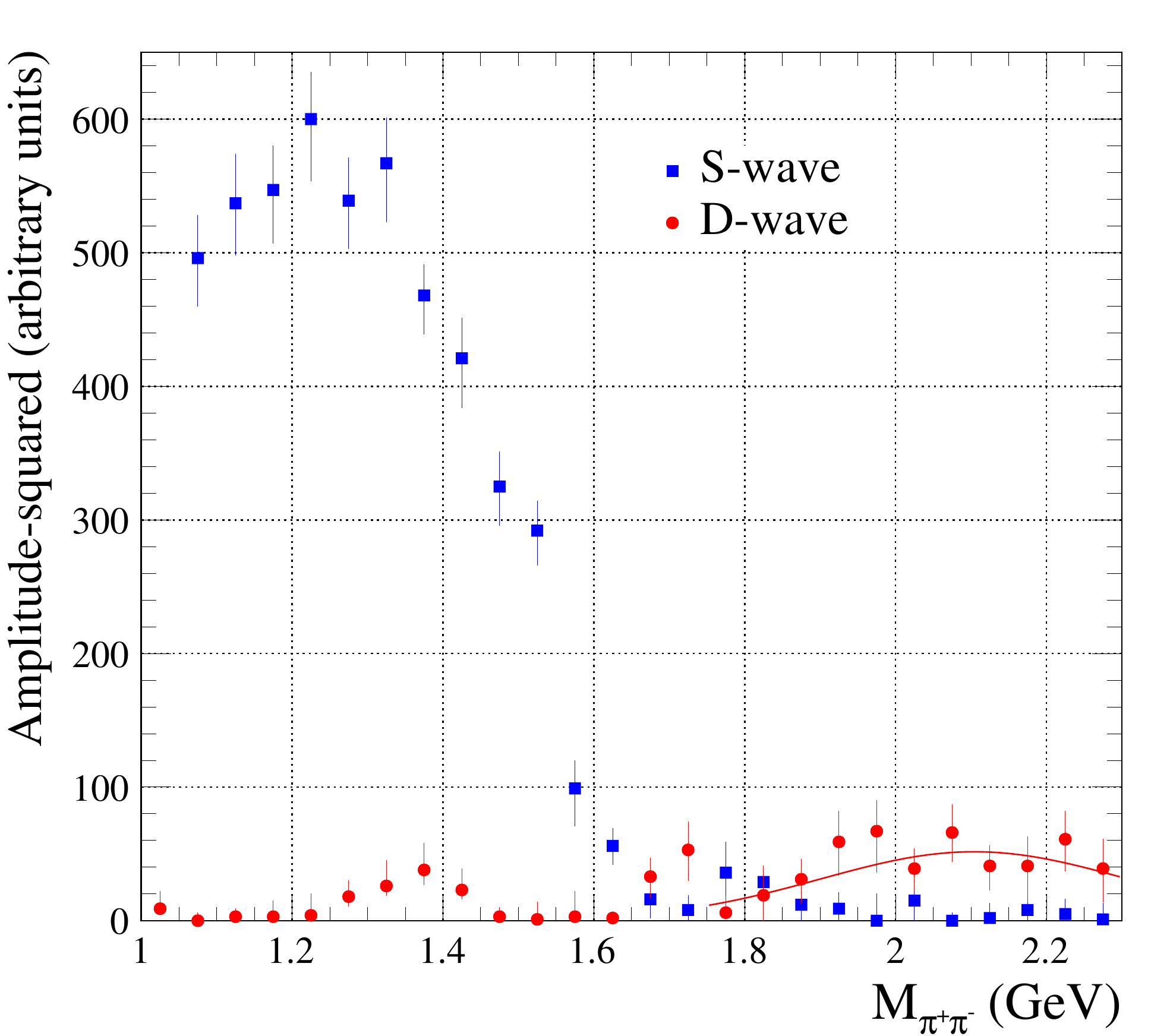}
}
\caption{Extraction of tensor-Pomeron parameters from a Gaussian fit to the exclusive $\pi^\pm$ Axial Field Spectrometer data: mean mass value $\left<M_{\pi^+\pi^-}\right>=2.10$~GeV and width $\Delta=\pm0.68$~GeV.}
\label{fig:fig1}
\end{figure}

The partial wave analysis of the {\sc afs} exclusive $\pi^\pm$ data~\cite{ref:PeterCesilThesis}, performed in terms of a fit with a model with S-wave and D-wave amplitudes as a function of the di-pion mass up to 2.3 GeV, leads to the results presented in Fig.~\ref{fig:fig1}. 

\noindent The D-wave dominates at masses above $\sim 2$~GeV, and according to the presumed interpretation in Ref.~\cite{ref:PeterCesilThesis} it corresponds to a spin-2 tensor glueball of mass M$_{\rm tgb}$. A Gaussian fit to this enhancement yields M$_{\rm tgb}=2.10\pm 0.68$~GeV. Identifying M$_{\rm tgb}^2$ with the saturated glueball-like enhancement of the {\sc mbr}-model parameter $s_0$ (see Eq.~\ref{eqTOT}) yields $s_0=4.42\pm 0.34$~GeV$^2$. Using this value in Eq.~\ref{eqTOT}, we predicted for  $\sigma_{\rm tot}$, $\sigma_{\rm el}$, and $\sigma_{\rm inel}$  at 13~TeV cross sections of $103.7\pm1.9$~mb, $30.2\pm0.8$~mb, and $73.5\pm1.3$~mb, respectively. The {\sc atlas}- and {\sc totem}-measured cross sections at $\sqrt s=7$ and 8 TeV~\cite{ref:atlas} \cite{ref:totem7TeV} \cite{ref:totem8TeV} are shown in Table~\ref{table} along with the {\sc mbr} predictions. The measurements are in good agreement with the predictions. Also shown is a recent measurement of the total inelastic cross section by {\sc atlas} at $\sqrt s=13$~TeV~\cite{ref:atlas13TeV}, $\sigma_{\rm inel}=73.1\pm0.9\;(\rm exp)\;\pm\rm 3.8\;{(\rm extr)}\pm6.6\;{(\rm lumi)}$~mb, which is in excellent agreement with the {\sc mbr} prediction.

\begin{table}[h!]
\caption{The total, elastic, and total inelastic MBR cross-section predictions (in mb) at $\sqrt s=$ 7, 8 and 13 TeV compared to measurements at the {\sc lhc} by {\sc totem} and {\sc atlas}. The tensor-Pomeron-based prediction of $\sigma_{\rm inel}$ at $\sqrt s=13$~TeV agrees with the {\sc atlas} measurement of $\sigma_{\rm inel}$~(exp) at the $\sim 1$\% level.}         
\vglue 0.5em
\centerline{\includegraphics[width=1.0\textwidth]{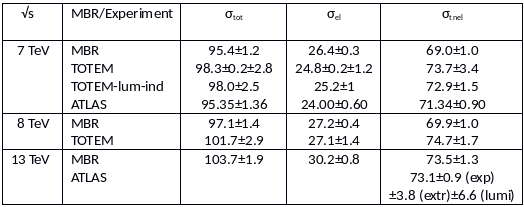}
}
\label{table}
\end{table}
It should be emphasized that the tensor-Pomeron hypothesis predics directly only the total cross section. As discussed above, the elastic cross section for $\sqrt{s} \le 1.8$ TeV is obtained from the global fit~\cite{ref:CMG_96}, while for $1.8<\sqrt s\leq 13$ TeV we use an extrapolation of the global-fit ratio of $\sigma_{\rm el}/\sigma_{\rm tot}$, which is slowly varying with $\sqrt s,$ multiplied by $\sigma_{tot}$.
The total inelastic cross section is calculated as the difference between the total and elastic. Thus, a measured lower $\sigma_{\rm el}$ would result in a higher $\sigma_{\rm inel}$. As seen in Table~\ref{table}, the {\sc mbr} $\sigma_{\rm el}$ is larger than the {\sc totem} and {\sc cms} measurements by $\sim 2$~mb at $\sqrt s$=7 TeV, which could imply a higher {\sc mbr} prediction for $\sigma_{\rm inel}$ at 13~TeV by $\sim 2$~mb as well.
This interplay between $\sigma_{\rm el}$ and $\sigma_{\rm inel}$ should be kept in mind as measurements of $\sigma_{\rm el}$ and $\sigma_{\rm tot}$ at $\sqrt s=13$~TeV become available in the near future.     

\section{Summary and conclusions}
We have presented predictions of the elastic scattering, total-inelastic, and total proton-proton cross sections at the {\sc lhc}, based on a Regge theory inspired tensor-Pomeron implementation of the RENORM model for hadronic diffraction. All measured cross sections are in good agreement within the experimental uncertainties of the data and the theoretical uncertainties of the model down to the $\sim $1\% level.
\section*{Acknowledgments}
Warmly acknowledged are Dr. Michael Albrow for providing a copy of the thesis of Peter C. Cesil, where the tensor-Pomeron data and analysis are discussed, and Dr. Robert Ciesielski for extracting the tensor-Pomeron parameters from the data and for many invaluable discussions.

\end{document}